\documentstyle [aps,psfig]{revtex}

\begin{document}
\draft
\title{Exact $U(1)$ symmetric cosmologies with local Mixmaster dynamics\thanks{
e-mail: berger@oakland.edu, vincent.moncrief@yale.edu}} 
\author{Beverly K. Berger}
\address{Department of Physics, Oakland University, Rochester, MI 48309 USA}
\author{Vincent Moncrief}
\address{Departments of Physics and Mathematics, Yale University, New Haven, CT
06520 USA}
\maketitle
\bigskip
\begin{abstract}
By applying a standard solution generating technique, we transform an arbitrary vacuum
Mixmaster solution on $S^3 \times {\bf R}$ to a new solution which is spatially
inhomogeneous. We thereby obtain a family of exact, spatially inhomogeneous, vacuum
spacetimes which exhibit Belinskii, Khalatnikov, and Lifshitz (BKL) oscillatory behavior.
The solutions are constructed explicitly by performing the transformations on numerically
generated, homogeneous Mixmaster solutions. Their behavior is found to be qualitatively like
that seen in previous numerical simulations of generic $U(1)$ symmetric cosmological
spacetimes on $T^3 \times {\bf R}$.
\end{abstract}

\section{Introduction}
Recent numerical studies \cite{berger98a} have provided strong evidence that 
$U(1)$-symmetric, vacuum spacetimes on $T^3\times 
{\bf R}$ generically
develop Mixmaster-like, oscillatory singularities of the type predicted
long ago by Belinskii, Khalatnikov and Lifschitz (BKL)
\cite{belinskii69a,belinskii69b,belinskii71a,belinskii71b,belinskii82}. These results
confirm numerically some of the most surprising features of the BKL prediction, namely that
nearby spatial points are effectively decoupled in their asymptotic metric evolution and
that the metric variables at each of these points evolve, at least qualitatively, like
those of a Mixmaster spacetime.

Several years ago B. Grubi\v si\' c and one of us (V.M.) \cite{grubisic94} made an 
analytical effort to generate some exact vacuum spacetimes which were
spatially inhomogeneous and which were expected to exhibit the sort of
oscillatory singularities which have since been seen in the  numerical
studies \cite{berger98a}.  That effort was not completed at the time since it was
not realized that several seemingly intractable integrals actually cancel
in the course of the calculations leaving only elementary computations to
be done.  We shall therefore complete that project here and use the
results to compare, in a more quantitative way, the numerical results
with some exact oscillatory singularities.

To generate new solutions having Mixmaster-like oscillations, we begin 
with the actual Mixmaster solutions and apply a standard solution
generating technique.  We choose one of the Killing fields shared by the
Mixmaster family and treat it as the generator of a spacelike $U(1)$
action on $S^3\times 
{\bf R}$, ignoring the presence of the other Killing
symmetries.  We compute the twist potential associated with the chosen
Killing field and reexpress the field equations, in a well-known way
\cite{moncrief86}, as a Kaluza-Klein reduced system on the base manifold
$S^2\times 
{\bf R}$ of the $S^1$-bundle $S^3\times 
{\bf R}\to
S^2\times 
{\bf R}$.  The field equations on the base take the form of
$2+1$ Einstein gravity coupled to a wave map whose target space is the
hyperbolic plane, $ 
{\bf H}^2$.  The isometry group of this latter space,
$SL(2, 
{\bf R})$, acts on the base fields in a natural way so as to
transform the given solution to a family of potentially inequivalent
solutions.

By a careful choice of the applied group element one can arrange that 
the transformed solution either lifts to the same bundle defined for the
original spacetime or perhaps to a different one (e.g., the trivial
bundle,$S^2\times S^1\times 
{\bf R}\to S^2\times 
{\bf R}$, or a ``squashed
sphere'', $S^3/ 
{\bf Z}_k\times 
{\bf R}\to S^2\times 
{\bf R}$).  Typically,
the new solutions will preserve only the Killing field that generates the
common $U(1)$ action and not preserve those Killing fields of the seed
solutions which fail to commute with the chosen $U(1)$ generator.  Thus
the new solutions are expected to be spatially inhomogeneous and yet to
exhibit Mixmaster-like oscillations inasmuch as their metrics are
parametrized by the same functions appearing in the Mixmaster seed
metrics themselves.

A previous application of this technique involved transforming an 
infinite dimensional family of ``generalized Taub-NUT'' spacetimes defined
on $S^3\times 
{\bf R}$, which have smooth Cauchy horizons at their
``singular'' boundaries, to a new family of curvature singular spacetimes
defined on $S^2\times S^1\times 
{\bf R}$ \cite{moncrief87}.  Because of the special
nature of the seed solutions in this case, the transformed solutions
developed only velocity dominated singularities and never exhibited
Mixmaster-like oscillations.  A new technique based upon expressing the
Einstein evolution equations in a so-called Fuchsian form seems capable
of significantly enlarging this set of rigorous, $U(1)$-symmetric,
curvature singular cosmological spacetimes but, so far, is also only
capable of yielding velocity dominated singularities \cite{kichenassamy98,isenberg00}.  So
far as we know the solutions presented for the first time here are the only
known exact inhomogeneous vacuum spacetimes which exhibit Mixmaster
oscillations.  Though only a finite dimensional family they presumably
display behavior representative of more general, $U(1)$-symmetric vacuum
spacetimes and thus warrant comparison with numerically produced
$U(1)$-solutions.  Making such a comparison is the second main aim, after
producing the solutions themselves, of this paper. As a byproduct of this
work, we also resolve a potential paradox that was pointed out in Ref.
\cite{grubisic94}.  There it was shown that every $U(1)$-symmetric vacuum
spacetime admits a certain gauge invariant conserved quantity which is
expressible purely locally in terms of the instantaneous Cauchy data for
that solution and serves as a Casimir invariant for the $SL(2, 
{\bf R})$ action.  For generic $U(1)$ solutions this quantity is known to be
non-trivial but, if non-trivial for the Mixmaster subfamily, would seem
to contradict the anticipated ``chaos'' of the Mixmaster dynamics
\cite{misner69,barrow82,cornish97b}.  The only sensible resolution, as was discussed in
Ref.~\cite{grubisic94}, is that the quantity actually vanishes on the Mixmaster subfamily. 
This we find to be the case by explicit calculation.

The inhomogeneity in our transformed solutions is produced, roughly 
speaking, by the fact that we choose to reduce with respect to a Killing
field which fails to commute with the remaining Killing fields of the seed
metric.  This is unavoidable with the generic Mixmaster solution but
special cases such as the Taub-NUT metrics allow for different
possibilities.  The additional Killing field admitted by Taub space
commutes with all the generators and is preserved upon reduction with
respect to one of these (non-abelian) generators.  The resulting
spacetime has therefore (at least) two commuting Killing fields and is thus
a special case of the so-called  Gowdy family of spacetimes.  By contrast
one could instead choose to reduce with respect to  the additional,
commutative Killling field but, in this case, all the symmetries are
preserved and one arrives, as was first shown by Geroch \cite{geroch71}, at only
the Kantowski-Sachs (i.,e. locally interior Schwarzschild) spacetime.

One might wonder if the ``new'' solutions we produce are really 
inhomogeneous at all or perhaps because of their expression in an unusual
gauge, are merely homogeneous solutions in disguise.  We shall use the
Gowdy transform of Taub space mentioned above, to show that this is not
the case---the new solutions are not in general globally homogeneous.

\section{Mixmaster spacetimes}
The Mixmaster spacetimes are spatially homogeneous vacuum metrics on 
$S^3\times 
{\bf R}$ whose line elements can be written
\begin{equation}
\label{eq1}
ds^2 = -N^2(t)dt^2 + A^2(t)(\hat\sigma^1)^2 + B^2(t)(\hat\sigma^2)^2 
+ C^2(t)(\hat\sigma^3)^2.
\end{equation}
Here the $\{ \hat\sigma^i\}$ are a global, analytic basis of one-forms on 
$S^3$ expressible in terms of the usual Euler angle coordinates $\{
x^1,x^2,x^3\} = \{ \theta,\varphi,\psi\}\in \{ [0,\pi), [0, 2\pi),
[0,4\pi)\}$ by 
\begin{eqnarray}
\label{eq2}
\hat\sigma^1 &=& \cos\varphi ~ d\theta + \sin \theta \sin \varphi ~ 
d\psi, \nonumber \\
\hat\sigma^2 &=& -\sin\varphi ~ d\theta + \sin\theta \cos\varphi ~d\psi, 
\nonumber \\
\hat\sigma^3 &=& d\varphi + \cos\theta ~ d\psi\,.
\end{eqnarray}
These forms, and therefore the above line element, are invariant with 
respect to the $U(1)$ action on $S^3$ generated by the Killing field
$\hat X_3 = \frac{\partial}{\partial\psi}$ whose orbits yield a Hopf
fibration of $S^3$, i.e. make $S^3$ into a principal fiber bundle over
$S^2$ with bundle projection given by
\begin{equation}
\label{eq3}
\pi_\psi :S^3\to S^2, (\theta, \varphi, \psi)\mapsto (\theta, \varphi).
\end{equation}

Of course the Mixmaster metrics are invariant with respect to a full 
$SU(2)$ action generated by Killing fields
\begin{eqnarray}
\label{eq4}
\hat X_1 &=& \cos \psi\frac{\partial}{\partial\theta} + \text{csc} ~ 
\theta\sin \psi \frac{\partial}{\partial\varphi} - \text{cot} ~ \theta
\sin \psi \frac{\partial}{\partial\psi}, \nonumber \\
\hat X_2 &=& -\sin \psi \frac{\partial}{\partial\theta} + \text{csc} ~ 
\theta \cos \psi \frac{\partial}{\partial\varphi} - \text{cot} ~ \theta
\cos \psi \frac{\partial}{\partial\psi}, \nonumber \\
\hat X_3 &=& \frac{\partial}{\partial\psi}
\end{eqnarray}
but, for the transformations we shall consider, only invariance with 
respect to $\hat X_3 = \frac{\partial}{\partial\psi}$ will in general
be preserved.

The equations of motion for the Mixmaster solutions are most simply 
expressed in a gauge for which $N = ABC$ where they take the form
\begin{eqnarray}
\label{eq5}
(\ln A^2),_{tt} &=& (B^2 - C^2)^2 - A^4, \nonumber \\
(\ln B^2),_{tt} &=& (C^2 - A^2)^2 - B^4, \nonumber \\
(\ln C^2),_{tt} &=& (A^2-B^2)^2 - C^4, 
\end{eqnarray} 
and are to be supplemented by the Hamiltonian constraint
\begin{eqnarray}
\label{eq6}
&&\frac{A_{,t}}{A} \frac{B_{,t}}{B} + \frac{A_{,t}}{A}\frac{C_{,t}}{C} 
+ \frac{B_{,t}}{B}\frac{C_{,t}}{C} + \nonumber \\
&&- \frac{1}{4} [A^4+B^4+C^4 - 2(A^2B^2 + B^2C^2 + A^2C^2)]\approx 0 .
\end{eqnarray}
In terms of the Misner anistropy variables $\alpha, \beta_{+},
\beta_{-}$,
\begin{eqnarray}
\label{eq7} 
A &=& e^{\alpha +\beta_{+}+\sqrt{3}~\beta_{-}}, \nonumber \\
B &=& e^{\alpha +\beta_{+}- \sqrt{3}~\beta_{-}}, \nonumber \\
C &=& e^{\alpha -2\beta_{+}},
\end{eqnarray}
and the chosen gauge condition is $N = e^{3\alpha}$. We now rewrite 
the line element in the $U(1)$-symmetric form developed in Refs.~\cite{moncrief86} and
\cite{moncrief90}.  Taking $\{ x^a\} = \{\theta,\varphi\}$ and noting that the shift vector
vanishes we express  $ds^2$ in the form
\begin{eqnarray}
\label{eq8}
&&ds^2 = e^{-2\gamma} \{ -\tilde N^2 dt^2 + \tilde g_{ab} dx^a dx^b\} \nonumber \\
&&+ e^{2\gamma} \{ d\psi + \cos \theta ~d\varphi + \beta_a dx^a\}^2
\end{eqnarray}
where $e^{2\gamma}$ is the scalar field $\frac{\partial}{\partial\psi}
\cdot\frac{\partial}{\partial\psi}$ given explicitly by
\begin{eqnarray}
\label{eq9}
e^{2\gamma} &=& A^2 \sin^2\theta \sin^2 \varphi + B^2 \sin^2 \theta 
\cos^2 \varphi \nonumber \\
&&+ C^2 \cos^2 \theta \,.
\end{eqnarray}
Since $\gamma$ is invariant with respect to the $U(1)$ action generated 
by $\hat X_3 = \frac{\partial}{\partial\psi}$
it induces a function on the quotient manifold $S^3\times 
{\bf R}/U(1)
\approx S^2\times 
{\bf R}$ which (with a slight abuse of notation) we
shall also designate by $\gamma$.  In a similar way one finds induced
upon the quotient manifold $S^2\times 
{\bf R}$ a Lorentzian metric
\begin{equation}
\label{eq10}
d\sigma^2 = -\tilde N^2 dt^2 + \tilde g_{ab} dx^a dx^b 
\end{equation}
and a one-form field
\begin{equation}
\label{eq11}
\mathop \beta \limits_\sim = \beta_a dx^a
\end{equation}
where $\{ a,b,\ldots\} = \{ 1,2\}$.  These forms are slightly 
specialized because of the vanishing of the shift vector field in
$ds^2$.  The most general $U(1)$-symmetric line element would yield
\begin{eqnarray}
\label{eq12}
d\sigma^2 &=& -\tilde N^2 dt^2 + \tilde g_{ab}(dx^a + \tilde N^adt)(dx^b
+\tilde N^b dt) \nonumber \\
\mathop \beta \limits_\sim &=& \beta_a dx^a + \beta_0 dt.
\end{eqnarray}

The explicit formulas for the $2+1$ Lorentzian metric $d\sigma^2$ and 
the one-form potential $
\mathop \beta \limits_\sim$ may be read off upon
expressing $ds^2$ in the form of Eq.~(\ref{eq1}).  One finds that
\begin{eqnarray}
\label{eq13}
\tilde N &=& Ne^\gamma,  \nonumber \\
\tilde g_{\theta\theta} &=& C^2 \cos^2\theta (A^2 \cos^2\varphi + B^2 
\sin^2\varphi) \nonumber \\
&&\qquad + A^2B^2 \sin^{2}\theta, \nonumber \\
\tilde g_{\varphi\varphi} &=& C^2 \sin^{2} \theta (A^2 \sin^{2} \varphi 
+ B^2 \cos^2 \varphi), \nonumber \\
\tilde g_{\theta\varphi} &=& - C^2 (A^2-B^2)\cos \varphi \sin\varphi 
\cos\theta \sin\theta, \nonumber \\
\beta_\theta &=& \frac{(A^{2}-B^{2})\cos\varphi \sin\varphi \sin\theta}
{e^{2\gamma}}, \nonumber \\
\beta_\varphi &=& (\frac{C^{2}\cos\theta}{e^{2\gamma}} - \cos \theta) , 
\end{eqnarray}
and computes, for example, that
\begin{equation}
\label{eq14}
\sqrt{\det ~^{(2)}\tilde g} = ABC \sin \theta e^\gamma.
\end{equation}

As in Refs.~\cite{moncrief86} and \cite{moncrief90} we introduce the momenta $\{\tilde p,
\tilde e^a, \tilde\pi^{ab}\}$ conjugate to $\{\gamma, \beta_a, \tilde g_{ab}\}$,
which, taken together, parameterize the full $3+1$ spatial metric
$g_{ij}$ and its conjugate momentum $\pi^{ij}$.  For the case of
vanishing shift the formulas relating the momentum variables $\{\tilde p,
\tilde e^a, \tilde\pi^{ab}\}$ to the metric variables $\{\gamma,
\beta_a,\tilde g_{ab}\}$ are given by
\begin{eqnarray}
\label{eq15}
\tilde p &=& \left(\frac{\sqrt{^{(2)}\tilde g}}{\tilde N} \right) 
4\gamma_{,t}, \nonumber \\
\tilde e^a &=& \left(\frac{\sqrt{^{(2)}\tilde g}}{\tilde N} \right) e^{4\gamma}
\tilde g^{ab} \beta_{b,t}, \nonumber \\
\tilde\pi^{ab} &=& \left(\frac{\sqrt{^{(2)}\tilde g}}{\tilde N} \right) \frac{1}{2} 
(\tilde g^{ac} \tilde g^{bd} - \tilde g^{ab} \tilde g^{cd}) \tilde g_{cd,t} \,.
\end{eqnarray}
For the Mixmaster metrics one computes that
\begin{eqnarray}
\label{eq16}
&&\tilde e^\theta =\frac{2\sin^{2}\theta}{N}\{BCA_{,t}\sin\varphi 
\cos\varphi - ACB_{,t}\sin\varphi \cos\varphi\}, \nonumber \\
&&\tilde e^\varphi = \frac{2\sin\theta\cos\theta}{N}\{ ABC_{,t}-BCA_{,t}
\sin^{2}\varphi-ACB_{,t}\cos^{2}\varphi\} , \nonumber \\
&&\tilde p = \frac{4 ABC \sin\theta}{Ne^{2\gamma}} [AA_{,t} \sin^{2} 
\theta \sin^2\varphi + BB_{,t}\sin^2\theta \cos^2\varphi + CC_{,t}
\cos^2\theta].
\end{eqnarray}
These momenta (along with $\tilde\pi^{ab}$ which we shall not need 
explicitly) project to yield smooth tensor densities on the base manifold
$S^2\times 
{\bf R}$ and one easily verifies that $\tilde e^a_{,a} = 0$
which is one of the components of the $3+1$ momentum constraint.

Note that the $U(1)$ connection one-form on $S^3\times 
{\bf R}$ given by
\begin{equation}
\label{eq17}
\mathop \lambda \limits_\sim := \hat\omega^3 + 
\mathop \beta \limits_\sim = 
d\psi + \cos \theta d\varphi + \beta_a dx^a
\end{equation}
does not project to yield a one-form on the base but that the difference 
between this connection one form and the reference one form
$\hat\omega^3:= d\psi + \cos \theta d\varphi$ does yield a one-form
(namely $
\mathop \beta \limits_\sim = \beta_a dx^a$) which projects to the
base.  Even though $
\mathop \lambda \limits_\sim$ itself does not project to
the base, its exterior derivative $
\mathop {d\lambda} \limits_\sim $ (i.e., the
curvature of the connection $
\mathop \lambda \limits_\sim$) does project.
Pulling back the induced two-form to a $t$ = constant slice of the
base manifold and computing its dual, one gets a scalar density $\tilde r$
defined by
\begin{eqnarray}
\label{eq18}
\tilde r &=& \in^{ab} \lambda_{a,b} \nonumber \\
&=& \in^{ab} \beta_{a,b} + \sin ~ \theta
\end{eqnarray}
whose explicit form is
\begin{eqnarray}
\label{eq19}
\tilde r &=& \frac{(A^2-B^2)}{(e^{2\gamma})^{2}} \sin^3\theta [B^2\cos^2
\varphi - A^2\sin^2\varphi] \nonumber \\
&&+ \frac{C^2\sin \theta}{(e^{2\gamma})^{2}} [A^2+B^2-C^2 +\sin^2\theta 
(C^2-A^2 \cos^2 \varphi -B^2 \sin^2\varphi)].
\end{eqnarray}
One computes on an arbitrary $t = $ constant slice of the base 
manifold, that
\begin{equation}
\label{eq20}
\hat B := \int_{S^{2}} \tilde r = 4\pi.
\end{equation}
The value $4\pi$ reflects the particular bundle $S^3\times 
{\bf R}\to 
S^2\times 
{\bf R}$ under study and would be the same for any $U(1)$-symmeteric 
metric defined on this bundle.

Taking into account the equation $\tilde e^a_{,a} = 0$ satisfied by 
$\tilde e^a$ and the fact that $S^2\times 
{\bf R}$
admits no non-trivial harmonic one forms, we now introduce the 
``twist potential'' function $\omega$ (a salar field on $S^2\times 
{\bf R}$) by imposing
\begin{eqnarray}
\label{eq21}
\tilde e^a &=& ~ \in^{ab} \omega_{,b}, \nonumber \\
\tilde r &=& \frac{\sqrt{^{(2)}\tilde g}}{\tilde N} e^{-4\gamma} 
\omega_{,t} \, .
\end{eqnarray}
These equations are self consistent and yield the solution
\begin{eqnarray}
\label{eq22}
\omega &=& \frac{\sin^2\theta}{N}\{ -ABC_{,t} +BCA_{,t} \sin^2\varphi 
+ACB_{,t} \cos^2\varphi \} \nonumber \\
&&+ k(t) 
\end{eqnarray}
which is unique up to the additive constant $k_0 := k(t_0)$ where $k(t)$ is the function defined by
\begin{equation}
\label{eq23}
k(t) = k(t_0) + \int^t_{t_{0}} dt' \left( \frac{N}{ABC}\right) C^2(A^2
+B^2-C^2).
\end{equation}
As discussed in Refs.~\cite{moncrief86} and \cite{moncrief90} the fields $\{ \gamma,\omega, 
d\sigma^2= -\tilde N^2 dt^2 + \tilde g_{ab} dx^a dx^b\}$ induced upon
the base manifold $S^2\times 
{\bf R}$ satisfy a $2+1$ dimensional system
of Einstein-wave map equations for which the target space of the wave map
is hyperbolic two-space (endowed with global coordinates
$\{\gamma,\omega\}$ and the natural metric $dh^2 = 4 d\gamma^2 +
e^{-4\gamma} d\omega^2$).  As a consequence of the $SL(2, 
{\bf R})$
isometry group of this target space the Einstein-wave map system admits
three independent constants of the motion which serve as the Hamiltonian
generators of the action of $SL(2, 
{\bf R})$ on the phase space of fields
$\{\gamma, \tilde p, \omega, \tilde r,\tilde g_{ab},\tilde\pi^{ab}\}$. 
These conserved quantities are given explicitly by the integrals
\begin{eqnarray}
\label{eq24}
\hat A &:=& \int_{S^{2}} (2\omega\tilde r + \tilde p) ,\nonumber \\
\hat B &:=& \int_{S^{2}} \tilde r ,\nonumber \\
\hat C &:=& \int_{S^{2}} (\tilde r (e^{4\gamma} - \omega^2) - \tilde p
\omega),
\end{eqnarray}
and we have already noticed that $\hat B = 4\pi$ for the Mixmaster 
spacetimes in particular.  In fact $\hat B$ would take this same value
for any $U(1)$-symmetric vacuum metric on $S^3\times 
{\bf R}$ but for
other $S^1$ bundles over the same base the value would (as discussed in
Refs.~\cite{moncrief86} and \cite{moncrief90}) be modified to $\hat B = 4\pi n$ where $n$ is
an integer determining the Chern class of the bundle.  In particular, for
solutions on the trivial bundle $S^2\times S^1\times 
{\bf R}$, $n$ would
vanish whereas if $n=2,3,\ldots ,$ the bundle would correspond to
various ``squashed spheres'' rather that a true $S^3$.

Note that, in view of the integral expression for $k(t)$ arising in the 
formula for $\omega$, both $\hat A$ and $\hat C$ are non-local in time. 
The same feature occurs in more general $U(1)$ symmetric solutions but
this non-locality cancels from the Casimir invariant
\begin{equation}
\label{eq25}
\hat K := \hat A^2 + 4\hat B\hat C
\end{equation}
which, however, vanishes identically for the Mixmaster family of 
solutions (though not in general).  The vanishing of $\hat K$ resolves a
potential mystery pointed out in Ref.~\cite{grubisic94} whereby a non-vanishing, local,
constant of the motion for Mixmaster metrics would seem to contradict
their empirically observed ``chaotic'' properties.

More specifically one finds, for the Mixmaster metrics, that
\begin{eqnarray}
\label{eq26}
\hat A &=& 8\pi k(t) + 8\pi \left(\frac{ABC}{N}\right)\left(\frac{A_{,t}}
{A} + \frac{B_{,t}}{B}\right), \nonumber \\
\hat B &=& 4\pi, \nonumber \\
\hat C &=& -\frac{\hat A^{2}}{16\pi} ,
\end{eqnarray}
so that $\hat K = 0$.  The non-locality of $\hat A$ and $\hat C$ 
sidesteps any conflict with the observed ``chaos'' in Mixmaster solutions
since, in fact, any Hamiltonian system will admit such non-local
constants of the motion.  To see this (even for a chaotic system) simply
time integrate Hamilton's equations and express the initial values of the
canonical variables in terms of time integrals of their driving
``forces.''

\section{The new solutions}
To generate new solutions of Einstein's equations from a given one (such 
as a Mixmaster solution) we choose an element $g\in SL(2, 
{\bf R})$,
\begin{equation}
\label{eq27}
g=\left( {\matrix{a&b\cr c&d\cr
}} \right)
,\qquad  ad - bc = 1
\end{equation}
and transform the fields $\{\gamma,\omega,\tilde p,\tilde r\}$ according 
to
\begin{eqnarray}
\label{eq28}
e^{2\gamma_{g}} &=& \frac{e^{2\gamma}}{[c^{2}(\omega^{2}+e^{4\gamma})
+2cd\omega +d^{2}]}, \nonumber \\
\omega_g &=& \frac{ac(\omega^{2}+e^{4\gamma})+(ad+bc)\omega + bd}{[c^{2}
(\omega^{2}+e^{4\gamma})+2cd\omega +d^{2}]} , \nonumber \\
\tilde p_g &=&\frac{\{\tilde p[c^2(\omega^2-e^{4\gamma})+2cd\omega 
+d^{2}]-\tilde r[4e^{4\gamma}(cd+\omega
c^{2})]\}}{[c^{2}(\omega^{2}+e^{4\gamma})+2cd\omega +d^{2}]}, \nonumber \\
\tilde r_g &=& \tilde p(c^2\omega + cd) + \tilde r[d^2+c^2(\omega^2
-e^{4\gamma})+2cd\omega],
\end{eqnarray}
while leaving $\{\tilde g_{ab}, \tilde\pi^{ab}, \tilde N, \tilde N^a\}$ 
invariant.  The induced transformation of the conserved quantities $\hat
A, \hat B, \hat C$ (by the so-called co-adjoint action of $SL(2, 
{\bf R})$)is found to be \cite{moncrief87}
\begin{eqnarray}
\label{eq29}
\hat A_g &=& (ad + bc) \hat A + 2bd \hat B - 2ac \hat C , \nonumber \\
\hat B_g &=& d^2 \hat B - c^2 \hat C + cd \hat A, \nonumber \\
\hat C_g &=& a^2 \hat C - b^2\hat B - ab \hat A, \nonumber \\
\hat K_g &=& \hat K = (\hat A_g)^2 + 4 \hat B_g \hat C_g\,.
\end{eqnarray}

To avoid a trivial transformation we shall require that $c$ be non-zero 
and, to ensure that the transformed solution lifts to an $S^1$ bundle
over$S^2\times 
{\bf R}$, we shall demand that
\begin{equation}
\label{eq30}
\hat B_g = 4\pi n, n = 0,1,2,\ldots  ~.
\end{equation}
Defining
\begin{equation}
\label{eq31}
\ell (t) := \frac{ABC}{N} \left(\frac{A,_{t}}{A} + \frac{B,_{t}}{B}
\right)
\end{equation}
we see from Eq.~(\ref{eq27}) that
\begin{eqnarray}
\label{eq32}
\hat A &=& 8\pi k(t) + 8\pi \ell(t) = 8\pi k(t_0) + 8\pi \ell(t_0) 
=8\pi (k_0+\ell_0), \nonumber \\
\hat B &=& 4\pi , \nonumber \\
\hat C &=& - 4\pi (k_0 +\ell_0)^2 \,.f
\end{eqnarray}
Setting $\hat B_g = 4\pi n, n=0,1,2,\ldots$ gives the restriction
\begin{equation}
\label{eq33}
[d + c(k_0 + \ell_0)]^2 = n\geq 0
\end{equation}
or, equivalently,
\begin{equation}
\label{eq34}
d + c(k_0 + \ell_0) = \pm n^{1/2}
\end{equation}
which can always be solved for $k_0$ since $c\ne 0$.

Exploiting the fact that $\hat A$, hence also $k(t) + \ell (t)$, is 
conserved one finds that the integral occurring in the formula for $k(t)$
can be expressed as
\begin{equation}
\label{eq35}
\int^t_{t_{0}} dt'\left(\frac{N}{ABC}\right) C^2(A^2+B^2-C^2) = -(\ell 
(t)-\ell_0)
\end{equation}
(this is also easily verified upon differentiation by using the equations
of motion (\ref{eq5})).  Using this result, one can easily show that
\begin{eqnarray}
\label{eq36}
&&(c\omega + d) = \pm n^{1/2} \nonumber \\
&&+ c ~~\frac{\sin^2\theta}{N} [-ABC_{,t} + BCA_{,t} \sin^{2}\varphi + 
ACB_{,t}\cos^2\varphi] \nonumber \\
&&- c ~~\frac{ABC}{N} \left(\frac{A_{,t}}{A} + \frac{B_{,t}}{B}\right) .
\end{eqnarray}
With this and Eq.~(\ref{eq9}) for $e^{2\gamma}$ one easily evaluates the 
transformed field variables $\{ e^{2\gamma_{g}}, \omega_g, \tilde p_g,
\tilde r_g\}$ using Eq.~(\ref{eq28}).  The new spacetime metric thus takes the
form
\begin{eqnarray}
\label{eq37}
ds^2_g &=& e^{-2\gamma_{g}} \{-\tilde N^2 dt^2 + \tilde g_{ab} dx^a 
dx^b\} \nonumber \\
&&+ e^{2\gamma_{g}} \{ d\psi + n \cos \theta d\varphi + \beta_{(g)a} 
dx^a\}^2
\end{eqnarray}
where however, $\beta_{(g)a}$ remains to be computed.  As discussed in 
Ref.~\cite{moncrief86}, $\beta_{(g)a}$ can be expanded (via the Hodge decomposition
for a one-form on $S^2$) as
\begin{equation}
\label{eq38}
\beta_{(g)a} = \left(\tilde g_{ac} \frac{\in^{cd}}{\tilde\mu_{g}}
\right) \eta_{,d} + \delta_{,a}
\end{equation}
where $\tilde\mu_g = \sqrt{^{(2)}\tilde g}$ and $\eta$ and $\delta$ 
are suitable functions defined on $S^2$.  The equation for $\beta_{(g)a} ~
dx^a$ is
\begin{equation}
\label{eq39}
\tilde r_g = \in^{ab} \beta_{(g)a,b} + n \sin ~\theta
\end{equation}
which, upon substitution of the decomposition (\ref{eq38}), becomes
\begin{equation}
\label{eq40}
\tilde r_g - n \sin ~\theta = (\tilde\mu_g \tilde g^{ab} 
\eta_{,a})_{,b}
\end{equation}
a Poisson equation for $\eta$ for which the necessary and sufficient 
integrability condition is ensured by Eq.~(\ref{eq30}).  This uniquely determines
$\eta$, at fixed $t$, up to an arbitrary additive constant and leaves
$\delta$ arbitrary.  The presence of $\delta$ reflects the freedom to
make an arbitrary coordinate transformation of the form $\psi\to\psi +
\delta$ without affecting the $U(1)$-form of the spacetime metric.

The time development of $\beta_{(g)a}$ can now be obtained by integrating
the (zero shift) evolution equation
\begin{eqnarray}
\label{eq41}
\beta_{(g)a,t} &=& \left(\frac{\tilde N}{\tilde\mu_{g}}\right) ~ 
e^{-4\gamma_{g}} ~ \tilde g_{ab} ~e^b_{(g)} \nonumber \\
&=& \left(\frac{\tilde N}{\tilde\mu_{g}}\right) ~ e^{-4\gamma_{g}} ~ 
\tilde g_{ab} \in^{bc} ~ \omega_{g,c}
\end{eqnarray}
with $\gamma_g, \omega_g$ determined as above.

Equations (\ref{eq39}) and (\ref{eq41}) are consistent with each other by virtue of the 
Hamilton equations satisfied by $\tilde r_g, \omega_g$.  Note that
whereas we have used the actual metric $\tilde g_{ab}$ in defining a Hodge
decomposition of $\beta_{(g)a}$, any smooth metric on $S^2$ could have
been used instead.  Furthermore one could have used Eq.~(\ref{eq40}) to determine
$\eta$ at an arbitrary time and then adjusted the time dependence of
$\delta$ to impose the zero shift condition which is implicit in Eq.~(\ref{eq41}).  In either
case the new metric (\ref{eq37}) will satisfy the vacuum field equations on the chosen $S^1$
bundle over $S^2\times 
{\bf R}$.

One might still wonder how we know that the transformed solutions are 
genuinely inhomogeneous.  Could they not be merely homogeneous solutions
disguised through the choice of a time slicing that is not adapted to the
(hypothetical) homogeneity?  To show that this is not the case, in
general,we shall examine a special case for which the transformed
solution has a hypersurface of time symmetry at $t=t_0$, i.e., has
$K^{(g)}_{ij}\mid_{t=t_{0}} = 0$.  To arrange this, we choose the
seed solution to have this property and make a careful choice of
transformation parameters so that the desired feature is not destroyed by
the $SL(2, 
{\bf R})$ transformation.  We then show that the transformed
spatial metric $g^{(g)}_{ij} \mid_{t=t_{0}}$ is not homogeneous as it
would have to be for the resulting spacetime to have this property.  The
key point here is the fact that on any compact slice, having constant mean
curvature the first and second fundamental forms $\{ g^{(g)}_{ij},
K^{(g)}_{ij}\}\mid_{t=t_{0}}$ would both have to be homogeneous in
order that the spacetime have this property.

Consider a Mixmaster solution for which $\dot A(t_0) = \dot B(t_0) = 
\dot C(t_0) = 0$.  This spacetime has the $t=t_0$ slice as a surface
of time symmetry and, because of the Hamiltonian constraint, must
satisfy
\begin{eqnarray}
\label{eq42} 
&&C(t_0) = \pm (A(t_0)\pm B(t_0)), \nonumber \\ 
&&A(t_0), B(t_0),
C(t_0) > 0. 
\end{eqnarray}
To maintain this property we choose the trivial target bundle $S^2\times 
S^1\times 
{\bf R}$ by  taking $n=0$.  We further simplify the
computations by choosing $A(t_0) = B(t_0) > 0$ and $C(t_0) = 2A(t_0)$
and find that the transformed metric at $t=t_0$ satisfies
\begin{eqnarray}
\label{eq43}
\tilde g_{\theta\theta}\mid_{t=t_{0}} &=& A^4(t_0)(\sin^2 \theta + 4 
\cos^2\theta), \nonumber \\
\tilde g_{\theta\varphi}\mid_{t=t_{0}} &=& 0 , \nonumber \\
\tilde g_{\varphi\varphi}\mid_{t=t_{0}} &=& 4 A^4(t_0)\sin^2\theta, \nonumber \\
{\mathop \beta \limits_\sim}{}_{(g)}\mid_{t=t_{0}} &=& \beta_{(g)a} dx^a
\mid_{t=t_{0}} = -c^2 4A^4(t_0)\cos\theta \sin^2\theta d\varphi, \nonumber \\
e^{2\gamma_{g}}\mid_{t=t_{0}} &=& \frac{1}{c^{2} A^{2}(t_0)(\sin^2
\theta +4\cos^2\theta)}\,.
\end{eqnarray}
Thus the new spatial metric induced at $t=t_0$ on $S^2\times S^1$ is
\begin{eqnarray}
\label{eq44}
d\ell^2_{(g)}\mid_{t=t_{0}} &=& c^2 A^2(t_0)(\sin^2\theta + 4\cos^2
\theta)\{4A^4(t_0)\sin^2\theta d\varphi^2 \nonumber \\ 
&&+   A^4(t_0)(\sin^2\theta +
4\cos^2\theta)d\theta^2\} \nonumber \\ 
&&+ \frac{1}{c^{2}A^{2}(t_{0})(\sin^{2}\theta
+4\cos^{2}\theta)} \{d\psi-4c^2A^4(t_0)\cos\theta \sin^2\theta d\varphi
\}^2.
\end{eqnarray}
A straightforward computation of $R_{ij}R^{ij}$ (the square of the Ricci 
tensor of this metric) proves that the resulting spacetime is not
homogeneous.  Indeed, the only vacuum homogeneous solution on $S^2\times
S^1\times 
{\bf R}$ is known to be the Kantowski-Sachs universe which does
not have a hypersurface of time symmetry.  It is possible to get the
Kantowski-Sachs metric upon transformation of a Taub metric (i.e., a
Mixmaster solution having $A(t) = B(t)$) but to do so one must reduce
with respect to the ``extra'' Killing field the Taub metric possesses,
$\frac{\partial}{\partial\varphi}$, rather than with respect to the
common Killing field $\frac{\partial}{\partial\psi}$ of the Mixmaster
family as we have done.  The extra Killing field
$\frac{\partial}{\partial\varphi}$ commutes with all the Killing
symmetries of the Taub solution and allows all of these symmetries to be
preserved upon reduction \cite{geroch71}.

The Taub metric used in the example above, is known explicitly but 
exhibits no $BKL$ type oscillations.  To see such oscillations in an
inhomogeneous setting, we combine a previously developed code for solving
the Mixmaster equations of motion with the transformations discussed
above.  Our results are discussed in the following section and compared
with results derived from a general $U(1)$-syummetric, vacuum Einstein
code.

\section{Numerical results}
Elsewhere we have shown that even the homogeneous Mixmaster model reproduces the local
behavior seen in generic $U(1)$-symmetric cosmologies \cite{berger00a}. From Eq.~(\ref{eq9}),
it is clear that $\gamma$ is dominated by the largest of the Mixmaster scale factors $A$,
$B$, or $C$. The local oscillations seen in $\gamma$ in the $U(1)$-symmetric models are
interpreted as follows:  Assume that the BKL approximate description of a homogeneous
Mixmaster model as a sequence of Kasner epochs is valid. In a given approximate Kasner
epoch assume  $A > B > C$ and that $A$ is increasing. Then $\gamma,_t > 0$ for $A,_t > 0$
while $B,_t$ and $C,_t$ are less than zero. The usual Mixmaster bounce changes the sign of
$A,_t$ and thus of $\gamma,_t$. However, after the bounce, either $B,_t$ (within an era) or
$C,_t$ (at the end of an era) becomes positive. When the growing scale factor surpasses the
decreasing $A$, $\gamma,_t$ will start to grow again since it will now track the new dominant
scale factor. A similar analysis indicates that the remaining ``dynamical'' variables,
$\omega$,
$\tilde p$, and $\tilde r$, depend on an order unity ratio of scale factors and thus do not
oscillate, as $\gamma$ does, between order unity and exponentially small values. (Here we
shall use ``order unity'' to mean some finite value which is not exponentially small.)

In our previous numerical simulations of generic $U(1)$-symmetric cosmologies on $T^3 \times
{\bf R}$, we noted that the oscillations in $\gamma$ could be interpreted as bounces off the
potentials $V_1 = {\textstyle {1 \over 2}} \tilde r^2 e^{4 \gamma}$ and $V_2 = {\textstyle {1
\over 2}} \tilde g\,\tilde g^{ab} e^{-4 \gamma} \omega,_a \omega,_b$. For a Mixmaster
solution, $V_1$ is exponentially small unless $e^{2 \gamma}$ is of order unity while $V_2$
is exponentially small unless the two largest scale factors are approximately equal to each
other \cite{berger00a}. This is clearly consistent with a presumption that the generic
models exhibit local Mixmaster dynamics.

To explore the nature of the new inhomogeneous $U(1)$-symmetric models, we note that the
transformed variables $\omega_g$, $\tilde p_g$, and $\tilde r_g$ will remain of order unity
(i.e. they will not oscillate between exponentially small and order unity values) because
the right hand sides of Eqs.~(\ref{eq28}b)-(\ref{eq28}d) are always of order unity. On the
other hand, $\gamma_g$ is dominated by the behavior of the oscillatory $\gamma$ since the
denominator on the right hand side of Eq.~(\ref{eq28}a) is always order unity while the
numerator oscillates.

To explore the differences between our new solution and the Mixmaster seed solution, we
construct the new solutions as follows:  First use the algorithm of Berger et al
\cite{berger96c} to obtain a numericallly generated Mixmaster model. This code is known to
solve the Mixmaster ODE's with machine-level precision and can follow hundreds of
bounces. The presumed stochastic properties of such a model imply that almost any Mixmaster
initial conditions will yield generic Mixmaster behavior. Thus, we need only consider a
single Mixmaster trajectory. Next, Eqs.~(\ref{eq9}), (\ref{eq16}), (\ref{eq19}),
(\ref{eq22}), (\ref{eq23}), (\ref{eq31}), and (\ref{eq35}) are used to numerically evaluate
$\gamma$, $\omega$, $\tilde p$, and $\tilde r$ from the numerically generated sequence of
values of the BKL scale factors and their time derivatives. Finally, for a representative
choice of the $SL(2, {\bf R})$ parameters and, e.g., $n=1$, the transformed variables
$\gamma_g$, $\omega_g$, $\tilde p_g$, and $\tilde r_g$ are computed using Eqs.~(\ref{eq28}). 

In Figures 1--3, we compare the
Mixmaster and transformed $\gamma$ and $\omega$ at a representative spatial point for
typical Mixmaster seed and set of $SL(2, {\bf R})$ parameters. Note that, in Fig.~1, the
original and transformed $\gamma$'s become indistinguishable after only a small number of
Mixmaster epochs. It is
clear that this will be so from Eq.~(\ref{eq9}) for
$\gamma$ and Eq.~(\ref{eq28}a) for the transformation. Since $\gamma$ and $\gamma_g$ are
found from the logarithm of Eqs.~(\ref{eq9}) and (\ref{eq28}a), both $\gamma$ and $\gamma_g$
will be approximately equal to the logarithm of the largest scale factor and depend only
logarithmically on the spatially dependent function associated with it. On a finer scale, in Fig.~2, the difference between the solutions
(especially near the ``bounce'' where $\gamma \approx \gamma_g \approx 0$) may be seen. 
On the other hand, as is seen in Fig.~3,
$\omega$ and $\omega_g$ are always of order unity and may easily be distinguished.

Figure 4 demonstrates the close link between Mixmaster dynamics and the oscillatory behavior
observed in our studies of generic $U(1)$-symmetric models and should be compared
to Figs.~2--6 in \cite{berger98a}. It shows the oscillations of $\gamma_g$ (or essentially
equivalently $\gamma$), $V_1$, and $V_2$ at a representative spatial point---reproducing
the behavior seen in our simulations of generic $U(1)$-symmetric cosmologies. Since we know
that these oscillations indicate local Mixmaster behavior in the new solutions, we can infer
that the observed oscillations in the generic models also indicate local Mixmaster dynamics.

Since $\gamma$ is the key variable in the $U(1)$-symmetric models and $\gamma \approx
\gamma_g$, one may then ask where these new $U(1)$-symmetric models differ from both
Mixmaster and generic $U(1)$-symmetric models. First, we emphasize that, except at special
values of the spatial coordinate angles, there are no qualitative differences attributable to
spatial topology. The Mixmaster spatial dependence of course represents a realization of the
Bianchi type IX symmetry. From Eq.~(\ref{eq9}), it is clear that three distinct spatial
patterns will appear in $\gamma$ (in the logarithm) depending on which scale factor
dominates. In Fig.~5, we compare the spatial dependence of $\gamma$ and $\gamma_g$ for 12
epochs of the seed Mixmaster solution. The epochs are arranged according to the dominant
scale factor. The numerical scale in each frame is chosen so that the average value of
$\gamma$ or $\gamma_g$ is the centroid. (If this were not done, no spatial dependence would
be visible.) From Eqs.~(\ref{eq9}) and (\ref{eq22}), $\gamma$ and $\omega$ have three
possible spatial dependences. The $SL(2, {\bf R})$ transformation of Eqs.~(\ref{eq28})
clearly mixes the spatial dependence of
$\gamma$ and $\omega$ to form $\gamma_g$ and $\omega_g$. In Fig.~5, we see the evolving
spatial dependence of $\gamma_g$. This is additional evidence that the new solutions are
spatially inhomogeneous.

In generic $U(1)$-symmetric models, one could qualitatively interpret the asymptotic approach
to the singularity as the evolution of a different Mixmaster model at every spatial point. In
particular, the Mixmaster epochs have spatially dependent durations---bounces at different
spatial points occur at different times. In contrast, our new solution is characterized
as is Mixmaster itself by spatially independent epoch durations since the new solution
bounces only when the seed solution does so. While one could modify the spacetime slicing to
yield spatially dependent epoch durations, one would expect to be able to detect the
difference between between a single underlying Mixmaster seed in the new solutions and a
continuum of approximate Mixmaster solutions in the generic case. 

\section*{Acknowledgments}
We would like to thank the Institute for Theoretical Physics at the University of
California / Santa Barbara for hospitality.  BKB would like to thank the Institute for
Geophysics and Planetary Physics of Lawrence Livermore National Laboratory for
hospitality. This work was supported in part
by National Science Foundation Grants PHY9732629, PHY9800103, PHY9973666, and PHY9407194.

\section*{Figure Captions}
\bigskip

Figure 1. Comparison of $\gamma$ and $\gamma_g$ at a typical spatial point. The Mixmaster
seed solution has initial values $\beta_+ =   -0.9847899998176387$,
$\beta_- =   0.09987655443789$,
$\Omega =  -8.00000000000000$,
$\dot \beta_+ =   -3.632980009876544$,
$\dot \beta_- =   4.58987654433567878$, and the Hamiltonian constraint (\ref{eq6}) solved
for $\dot \Omega$. The $SL(2, {\bf R})$ parameters are $a = 1$, $b = 1$, $c = 10000$, and $d
= 10001$.

\bigskip

Figure 2. Detail of the comparison of $\gamma$ and $\gamma_g$. To emphasize the approach of
$\gamma_g$ to $\gamma$, data from later in the simulation
of Fig.~1 are shown. The actual, saved data values are indicated by the $\times$ and
$+$ symbols.

\bigskip

Figure 3. Comparison of $\omega$ and $\omega_g$ for the same models as in Fig.~1. Note that
$\omega_g$ appears to decrease to zero. This is due to the fact that choice of $SL(2, {\bf
R})$ parameters causes $|\omega_g| \approx 10^{-4}$ if $|\omega| \approx 1$.

\bigskip

Figure 4. New solution as an inhomogeneous $U(1)$-symmetric cosmology. As in 
previous studies of generic $U(1)$-symmetric cosmologies, $\gamma_g$, $V_1$, and $V_2$ are
shown at a typical spatial point. 

\bigskip

Figure 5.  Evidence for the spatial inhomogeneity of the new solutions. The spatial
dependence of $\gamma$ and $\gamma_g$ is shown for the $(\cos \theta, \varphi)$ plane in a
series of side-by-side frames arranged in three separate panels. Each pair of frames shows
the spatial dependence of $\gamma$ and $\gamma_g$ respectively during an approximate Kasner
epoch of the seed Mixmaster solution. The panels are grouped according to the identity of the
dominant scale factor in the spatially homogeneous solution rather than sequentially.
According to Eq.~(\ref{eq9}),
$\gamma$ will have the spatial dependence $\ln(\sin \theta\,\sin \varphi)$, $\ln
(\sin \theta \, \cos \varphi)$, or $\ln (\cos \theta)$ depending on whether $A$, $B$, or $C$
respectively is dominant. In each of the three panels, the 4 left-hand frames reproduce one
of these three spatial dependences with, reading from left to right, $B$, $C$, or $A$
dominant. In each case, the accompanying right-hand frame represents the spatial dependence
of the corresponding $\gamma_g$ for that epoch. In every case, the numerical scales for
$\gamma$ and
$\gamma_g$ have been centered on their average values to enhance the visibility of the
spatial dependence.

\begin{figure}[bth]
\begin{center}
\makebox[4in]{\psfig{file=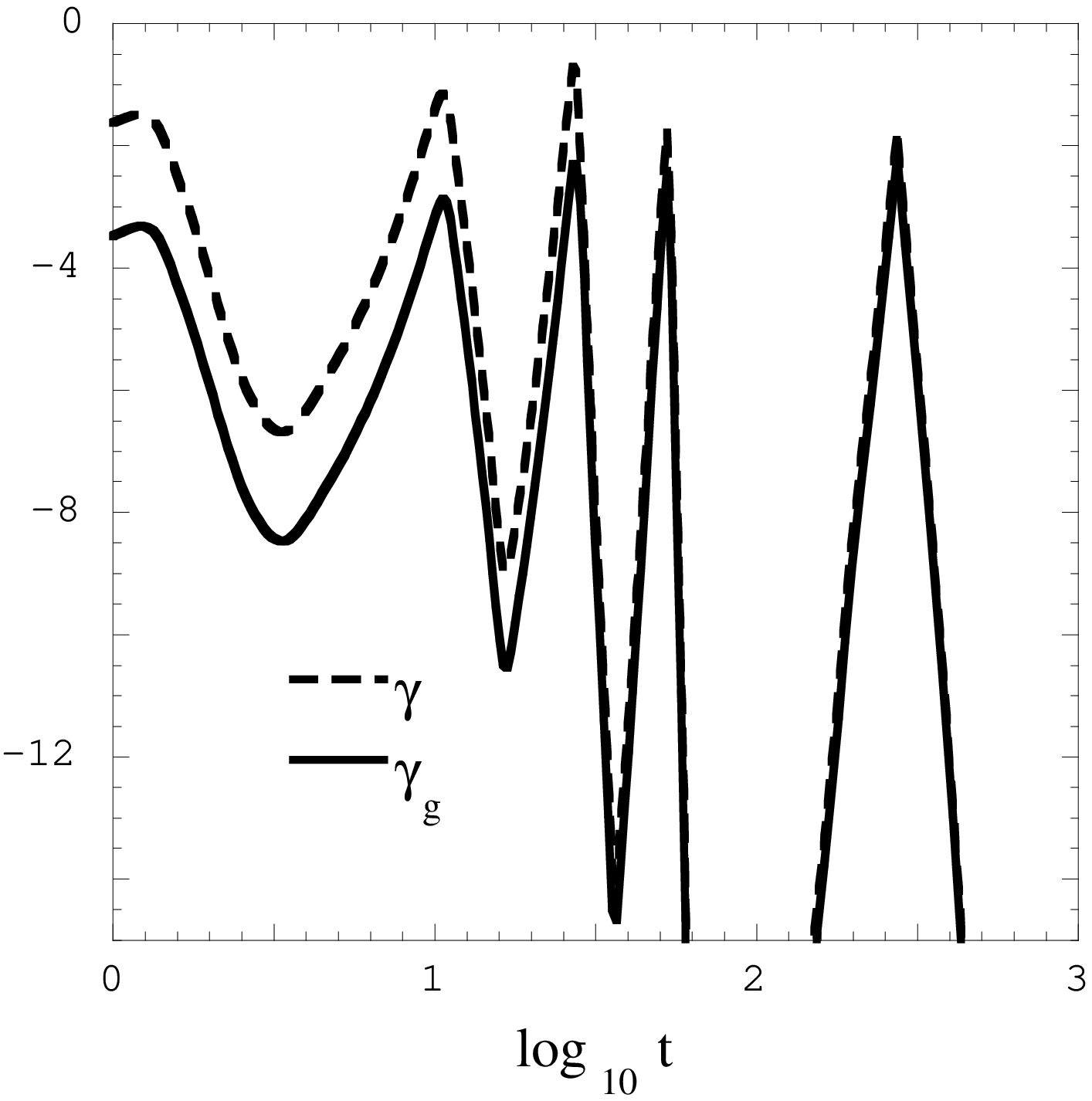,width=3.5in}}
\caption{}
\end{center}
\end{figure}
\begin{figure}[bth]
\begin{center}
\makebox[4in]{\psfig{file=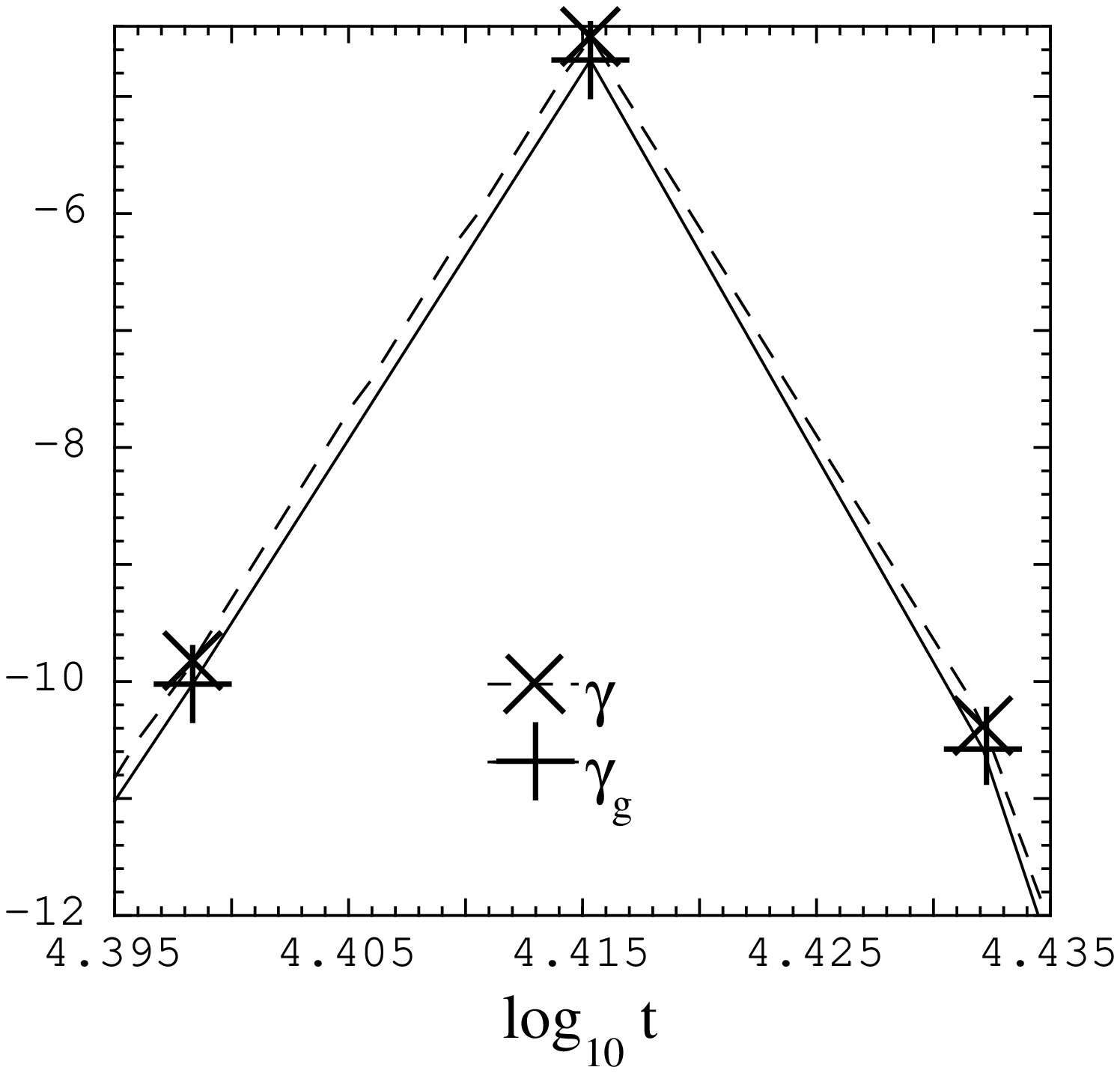,width=3.5in}}
\caption{}
\end{center}
\end{figure}
\begin{figure}[bth]
\begin{center}
\makebox[4in]{\psfig{file=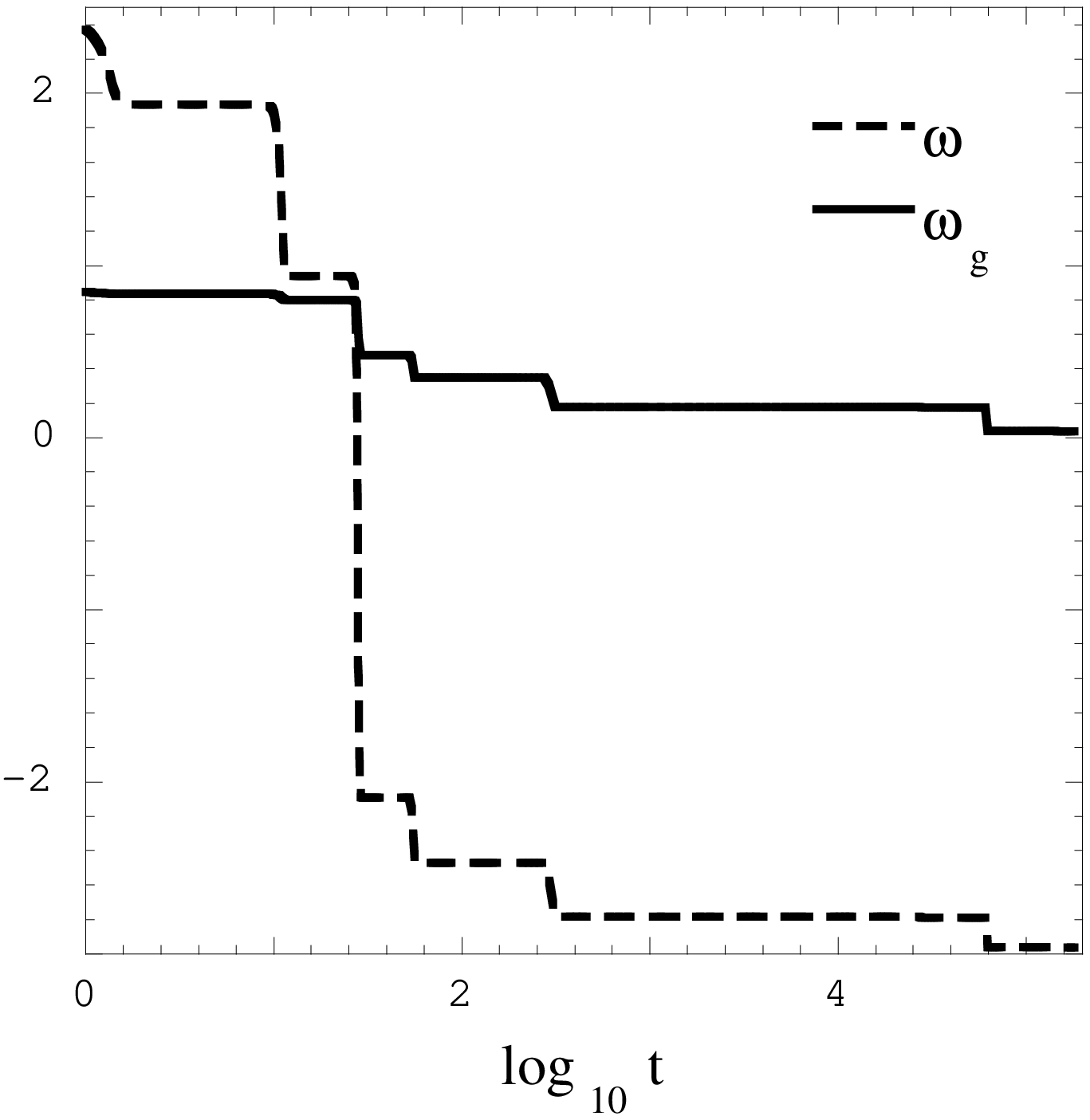,width=3.5in}}
\caption{}
\end{center}
\end{figure}
\begin{figure}[bth]
\begin{center}
\makebox[4in]{\psfig{file=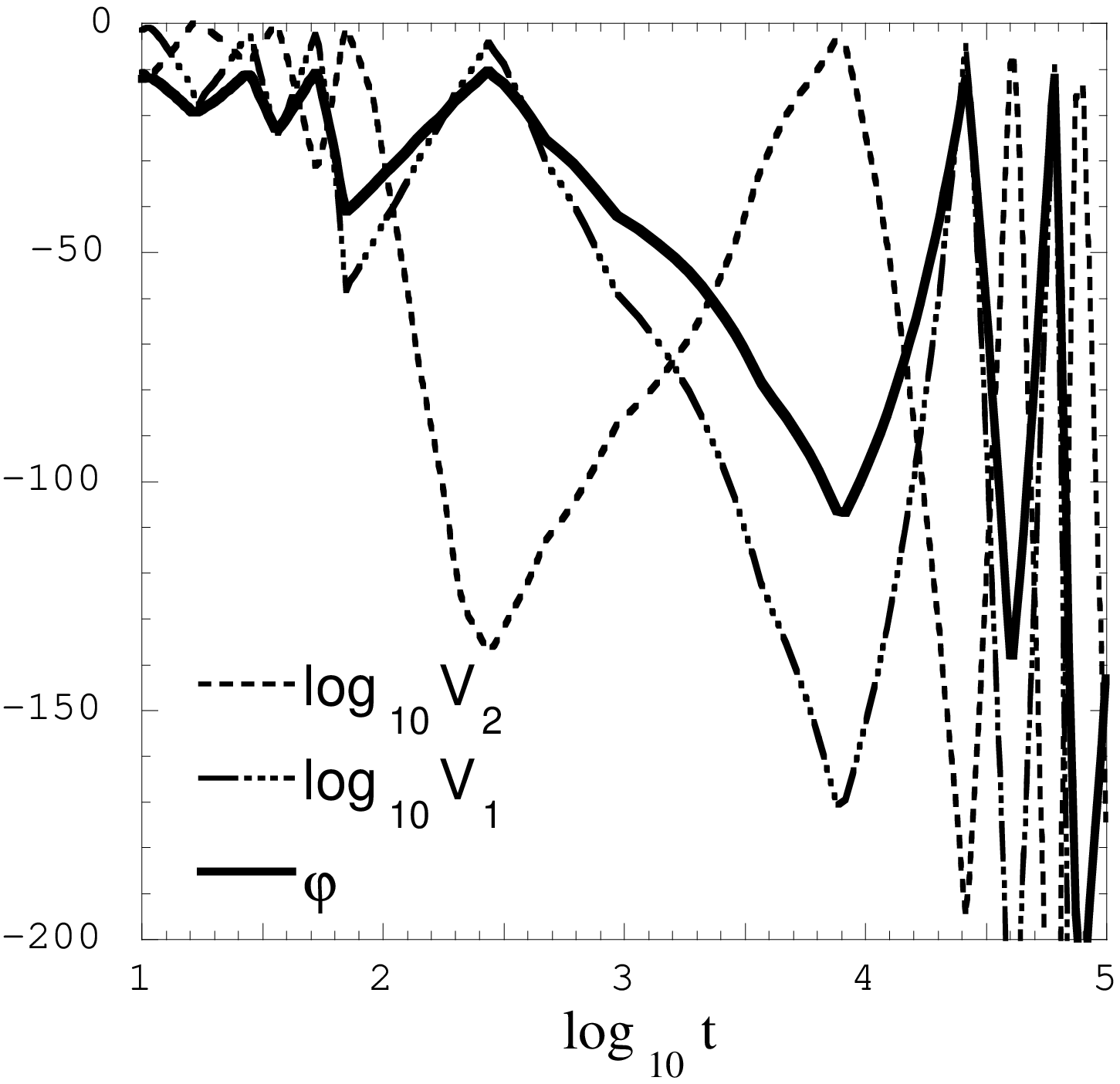,width=3.5in}}
\caption{}
\end{center}
\end{figure}
\begin{figure}[bth]
\begin{center}
\makebox[4in]{\psfig{file=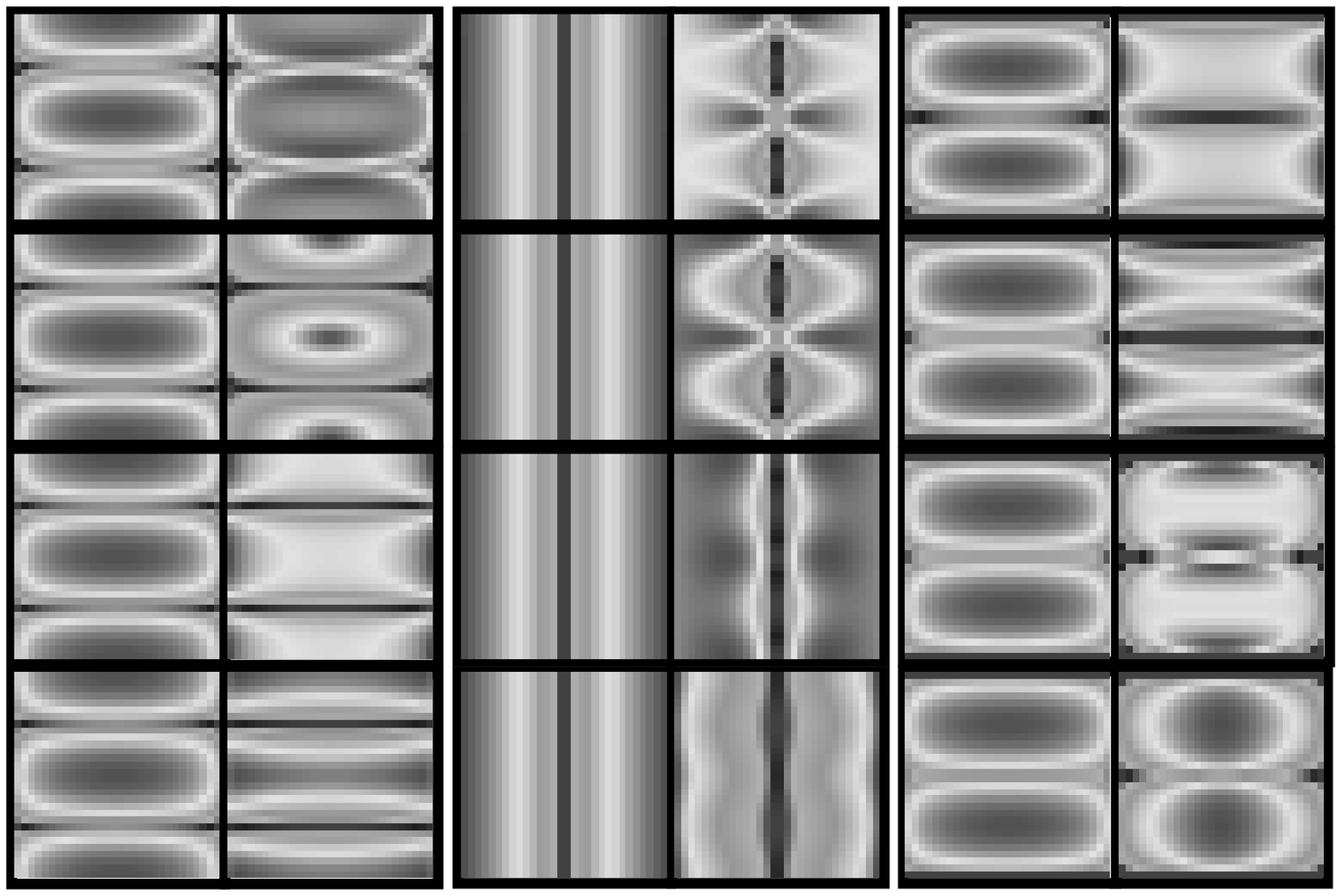,width=4.0in}}
\caption{}
\end{center}
\end{figure}

\end{document}